\documentclass[a4paper,11pt]{article}

\title{The Link between General Relativity and Shape Dynamics}
\author{\bf Henrique Gomes\footnote{\href{mailto:gomes.ha@gmail.com}{gomes.ha@gmail.com}}\\\it  Department of Physics,  University of California, Davis,   CA, 95616 \bigskip\\ \bf Tim Koslowski\footnote{\href{mailto:tkoslowski@perimeterinstitute.ca}{tkoslowski@perimeterinstitute.ca}}
\\\it Perimeter Institute for Theoretical Physics\\\it 31 Caroline Street, Waterloo, Ontario N2L 2Y5, Canada}

\usepackage{amssymb}
\usepackage{amsmath}
\usepackage[usenames,dvipsnames]{color}
\usepackage{hyperref}
\usepackage[left=1in,right=1in,top=1in,bottom=1in]{geometry}                  

\hypersetup{
    colorlinks=true,         
    linkcolor=blue,          
    citecolor=red,        
    urlcolor=Violet             
}

\let\oldmarginpar\marginpar
\renewcommand\marginpar[1]{\oldmarginpar{\color{red}\raggedright\scriptsize #1}}

\newcommand{\mean}[1]{\ensuremath{\lf\langle #1 \rt\rangle }}
\newtheorem{proposition}{Proposition}

\def\be{\begin{equation}}
\def\ee{\end{equation}}
\def\bea{\begin{eqnarray}}
\def\eea{\end{eqnarray}}

\def\lf {\ensuremath{\left}}
\def\rt {\ensuremath{\right}}

\def\hg{\mathcal H_{\textrm{gl}}}
\begin{document}

\maketitle
\begin{abstract}
  We define the concept of a linking theory and show how two equivalent gauge theories possessing different gauge symmetries generically arise from a linking theory. We show that under special circumstances a linking theory can be constructed from a given gauge theory through ``Kretchmannization'' of a given gauge theory, which becomes one of the two theories related by the linking theory. The other, so-called ``dual'' gauge theory, is then a gauge theory of the symmetry underlying the ``Kretschmannization''. We then prove the equivalence of General Relativity and Shape Dynamics, a theory with fixed foliation but spatial conformal invariance. This streamlines the rather complicated construction of this equivalence performed in \cite{Gomes:2010fh}. We use this streamlined argument to extend the result to General Relativity with asymptotically flat boundary conditions. The improved understanding of linking theories naturally leads to the Lagrangian formulation of Shape Dynamics, which allows us to partially relate the degrees of freedom.
\end{abstract}

\section{Introduction}

Gauge symmetries are a very important tool and a useful guiding principle for the construction of physical models, as the specification of degrees of freedom and symmetries is one of the first steps in defining a physical system. These gauge symmetries manifest themselves as sets of first class constraints in the classical Hamiltonian description.

In \cite{Gomes:2010fh} we took inspiration from \cite{barbour:bm_review,Barbou:CS_plus_V,Barbour:shape_dynamics} and used the St\"uckelberg mechanism \cite{stueckelberg} and methods from \cite{Dirac:CMC_fixing,York:york_method_prl,York:york_method_long} to construct equivalent gauge theories that have different sets of first class constraints, which turn out to generate different gauge symmetries. This unexpected result relies on the observation that in some cases one is able to find gauges for both systems such that the initial value problems and the equations of motion of both systems coincide, so the two systems do indeed have equivalent trajectories, despite their different gauge symmetries\footnote{A similar phenomenon occurs in the AdS/CFT correspondence \cite{Maldacena:ads_cft}, particularly as explored in \cite{Freidel:2008sh}.}. We applied this procedure to General Relativity on a compact manifold without boundary and found an equivalent theory that was a gauge theory of volume-preserving 3-dimensional conformal transformations, which we will call Shape Dynamics\footnote{The name Shape Dynamics indicates that our theory was inspired by Brabour's relational ideas, and in particular that it is a metric gauge theory of diffeomorphisms and local conformal transformations, implying that the physical degrees of freedom are given by the shape degrees of freedom of the spatial metric. We caution however that our particular theory is not the mechanical model \cite{1} that Brabour named ``Dynamics of pure shape'', nor is it equivalent to the notion of shape used by Anderson \cite{2,3} and Kendall \cite{4}.}.

However, using the St\"uckelberg construction\footnote{The St\"uckelberg construction is called ``Kretschmannization'' in the Relativity community. See \cite{Kuchar:1990qw} for a detailed discussion of Kretschmannization and parametrization of symmetries.} obscures some underlying structure that makes the equivalence possible. This structure is made most transparent by defining the key properties of a linking gauge theory that admits two compatible partial gauge fixings which yield the two equivalent gauge theories through partial phase space reduction. It is the purpose of this paper to bring this underlying structure to light and to investigate the relation of the Lagrangians of General Relativity and Shape Dynamics.

This paper is organized as follows: In section \ref{sec:EquivGT} we define linking gauge theories and show how pairs of equivalent gauge theories are constructed from them and show  how linking gauge theories can be constructed from a give gauge theory that has a certain property. We then apply this formalism in section \ref{sec:applyToGR} to General Relativity with asymptotically flat boundary conditions and construct the related Shape Dynamics theory. In section \ref{sec:Lagrangian}, we consider the Lagrangian version of Shape Dynamics.

\section{Equivalence of Gauge Theories}\label{sec:EquivGT}

This section is intended to show the mechanism that relates equivalent gauge theories in its generality. We show that whenever there is a linking gauge theory that links two gauge theories then these two gauge theories are equivalent. In such a case one should not consider a gauge theory by itself but the true structure requires the examination of both symmetry manifestations. After introducing linking theories in the first subsection, we give a general construction principle in the following subsection that can be used to construct a linking theory from a theory that possesses certain necessary properties, which are discussed in more detail in the final subsection.

\subsection{Linking Theories}

A gauge theory can be denoted by data $T=(\Gamma,\{.,.\},\{\chi_i\}_{i\in \mathcal I},\{\rho_j\}_{j\in \mathcal J})$, where $\Gamma$ denotes the phase space carrying the Poisson structure $\{.,.\}$, the set $\{\chi_i\}_{i\in\mathcal I}$ denotes first class constraints and the set $\{\rho_j\}_{j\in\mathcal J}$ denotes second class constraints\footnote{Second class constraints are at this point introduced solely for later notational convenience. We use DeWitt's condensed index notation. We will however later focus on the the case where the linking theory itself does not possess second class constraints.}. We study a class of theories with no explicit Hamiltonian; it can be included in the set of first class constraints as the constraint $H-\epsilon$ that enforces energy conservation. The initial value problem of $T$ is given by finding the space $\mathcal C=\{x\in \Gamma:\chi_i(x)=0\forall i\in\mathcal I\}\cap\{x\in \Gamma:\rho_j(x)=0\forall j\in\mathcal J\}$ and the canonical equations of motion are given by the Hamilton vector fields $v_H(\lambda_i)$ defined through the action on smooth phase space functions $f$ as
\begin{equation}
 v_H(f) = \{f,\sum_{i\in\mathcal I} \lambda_i \chi_i+\sum_{j\in \mathcal J} \mu_j \rho_j\},
\end{equation}
where the $\lambda_i$ are arbitrary Lagrange multipliers and the $\mu_j$ are fixed by the condition that $v_H$ is tangent to $\mathcal C$. Furthermore, one is able to impose (partial) gauge-fixing conditions $\{\sigma_i\}_{i\in \mathcal I^o}$, such that (some of) the Lagrange multipliers $\lambda_i$ are determined by the condition that $v_H$ is tangent to $\mathcal C_{gf}=\mathcal C \cap \{x\in \Gamma:\sigma_i(x)=0\forall i\in\mathcal I^o\}$. Hence, gauge-fixing conditions turn (some of) the first class constraints into second class constraints and turn the initial value problem into a gauge-fixed initial value problem $\mathcal C_{gf}$.

There can be a nontrivial physical equivalence between gauge theories, because physical quantities are gauge-invariant. To be precise, two gauge theories $T_1,T_2$ are {\bf  equivalent} if there is a (partial) gauge-fixing $\Sigma_1=\{\sigma^1_i=0\}_{i\in \mathcal I^o_1}$ of $T_1$ and another partial gauge fixing $\Sigma_2=\{\sigma^2_i=0\}_{i\in \mathcal I^o_2}$ of $T_2$, such that the initial value problems $C^1_{gf}=C^2_{gf}$ and the gauge-fixed Hamilton-vector fields coincide.

We define a {\bf general linking} gauge theory as a triple $L=(T_L,\Sigma_1,\Sigma_2)$, where
$$T_L=(\Gamma_{\mbox{\tiny Ex}},\{.,.\},\{\chi_i\}_{i\in\mathcal I},\{\rho_j\}_{j\in \mathcal J})$$
is a gauge theory as described before\footnote{The reason for subscript Ex (denoting that $\Gamma$ is extended) is that the phase space of the linking theory is larger than the phase space $\Gamma$ of the two equivalent gauge theories.} and $\Sigma_1=\{\sigma^1_k\}_{k\in \mathcal K}$ and $\Sigma_2=\{\sigma^2_l\}_{l\in\mathcal L}$ are two sets of partial gauge fixing conditions such that $\Sigma_1 \cup \Sigma_2$ is a (partial) gauge-fixing condition for $T_L$. Furthermore, we demand that we can split the set $\mathcal X=\{\chi_i\}_{i\in \mathcal I}$ of first class constraints into three disjoint subsets: $\mathcal X_1,\mathcal X_2$ and $\mathcal X_o$, where $\mathcal X_1$ is gauge fixed by $\Sigma_1$, $\mathcal X_2$ is gauge fixed by $\Sigma_2$ and $\mathcal X_o$ is neither gauge fixed by $\Sigma_1$ nor by $\Sigma_2$.

Given a linking gauge theory, we can trivially construct two equivalent gauge theories:  We see that
$T_1=(\Gamma_{\mbox{\tiny Ex}},\{.,.\},\mathcal X_o\cup \mathcal X_2,\{\rho_j\}_{j\in \mathcal J}\cup\Sigma_1\cup\mathcal X_1)$ and $T_2=(\Gamma_{\mbox{\tiny Ex}},\{.,.\},\mathcal X_o\cup \mathcal X_1,\{\rho_j\}_{j\in \mathcal J}\cup\Sigma_2\cup\mathcal X_2)$ are equivalent gauge theories, because both can be gauge-fixed to $(\Gamma_{\mbox{\tiny Ex}},\{.,.\},\mathcal X_o,\{\rho_j\}_{j\in \mathcal J}\cup\Sigma_1\cup\Sigma_2\cup\mathcal X_1\cup\mathcal X_2)$. 

To complete the construction we construct the Dirac-bracket and reduced phase space for $T_L^1$ and $T_L^2$. This is where we come to the most important property that transforms the general concept of linking theory into an interesting non-trivial tool. This happens in the {\bf special} case\footnote{The examples we will consider are all of this case.} where (1) the phase space $\Gamma_{\mbox{\tiny Ex}}$ is a direct product of a phase space $\Gamma$ and an extension that is coordinatized by a canonically conjugate pair $\{\phi_i,\pi_\phi^i\}_{i\in\mathcal I}$ and the set of first class constraints take the form
\begin{equation}\label{equ:special-constraints}
 \begin{array}{rcl}
  &\chi^1_i:=\phi_i - f_i \approx & 0\\
   &\chi^i_2:= \pi_\phi^i - g_i \approx & 0,
 \end{array}
\end{equation}
where $f_i, g_i$ are functions on $\Gamma$ for all $i\in \mathcal I$. To complete the definition of a {\bf special} linking theory we demand the special gauge-fixing conditions
\begin{equation}\label{equ:special-gaugefixing}
\Sigma_1= \phi_i= 0,\,\,\,\,\Sigma_2=\pi_\phi^i= 0
\end{equation}
for all $i\in \mathcal I$. We will construct special linking theories through ``Kretschmannization'' in the next subsection, where it turns out that the decisive point is the existence and form of the constraints $\chi^1_i$, while the form of the constraints $\chi_2^i$ follows from the construction principle.

Note that $\Sigma_1$ completely fixes the gauge of $\chi_2 $ to zero and vice-versa. The special form of the constraints and gauge fixing conditions allows us to perform the phase space reduction explicitly\footnote{Dirac attempted a gauge fixing of this sort \cite{Dirac:CMC_fixing}, taking advantage of its special properties, but he did not consider a linking theory of any sort and thus fell short of finding shape dynamics.}: For this we consider functions $F_r$ on $\Gamma_{\mbox{\tiny Ex}}$ that are independent of $\{\phi_i,\pi_\phi^i\}_{I\in\mathcal I}$, which are in one-to-one correspondence with functions on $\Gamma$, and construct their Dirac-bracket $\{.,.\}_D$ for the gauge-fixing $\phi_i\approx 0$:
\begin{equation}\label{Dirac bracket}
 \{F_1,F_2\}_D=\{F_1,F_2\}+\{F_1,\phi_i\}\{\pi_\phi^i-f^i,F_2\}-\{F_1,\pi_\phi^i-f^i\}\{\phi_i,F_2\}=\{F_1,F_2\},
\end{equation}
where Einstein summation over $i$ is assumed, and we used the facts that $\{\phi_i,g^j\}=0$ and that $\phi, \pi_\phi$ are canonically conjugate. The Dirac bracket thus reduces to the Poisson bracket on the reduced phase space $\Gamma \subset \Gamma_{\mbox{\tiny Ex}}$ and the remaining first class constraints are
\begin{equation}
 f_i \approx 0, \,\,\textrm{ for all }i\in \mathcal I.
\end{equation}
Performing the analogous phase space reduction for the gauge-fixing condition $\pi_\phi^i\approx 0$, we arrive at
\begin{proposition}\label{prop:equivalence}
  If there exists a gauge-theory on a phase space $\Gamma_{\mbox{\tiny Ex}}=\Gamma\times \tilde \Gamma$, with special constraints of the form (\ref{equ:special-constraints}) and thus special gauge fixing conditions of the form (\ref{equ:special-gaugefixing}) then $T_1=(\Gamma,\{.,.\},\{f_i\}_{i\in\mathcal I}\cup \mathcal X_o,\{\rho_j\}_{j\in \mathcal J})$ and $T_2=(\Gamma,\{.,.\},\{g_i\}_{i\in\mathcal I}\cup \mathcal X_o,\{\rho_j\}_{j\in \mathcal J})$ are equivalent gauge theories.
\end{proposition}
Note that this proposition only assumes that the constraint can be formally written in the form (\ref{equ:special-constraints}). However, any set of constraints that can in principle be solved for $\phi_i$ and $\pi^i_\phi$ at the respective gauge fixing surface as in (\ref{equ:special-constraints}) suffices for the construction of the phase space reduction.

\subsection{A Construction Principle for Linking Theories}\label{sec:constructionPrinciple}

We will now give a simple construction principle for special Linking Theories linking a known gauge theory to a desired gauge theory with different symmetry. This is closely related to the Lagrangian concept of best matching\cite{barbour:bm_review}; we will comment on the relation at the end of this subsection. For this purpose we consider elementary degrees of freedom $q_i$ whose dynamics is governed by an action $S[q]:=\int dt L(\dot q,q)$. If the dynamical system is consistent then the Legendre transform to the canonical system will yield first and second class constraints, however we neglect second class constraints in this subsection (as they will not appear in such a fashion in the model we will be studying) and assume to have a conjugate pair $(q_i,p^i)$ of canonical degrees of freedom that coordinatize our phase space $\Gamma$ and solely a first class system of constraints
\begin{equation}\label{equ:originalConstraints}
 \chi_\rho(q,p)\approx 0.
\end{equation}
In the Lagrangian picture, let us now extend our configuration space to include auxiliary degrees of freedom $\phi_\alpha$, but still with the Lagrangian $L(\dot q,q)$, which implies that the Legendre transform yields a phase space with an additional canonically conjugate pair $(\phi_\alpha,\pi^\alpha)$ and with additional first class constraints
\begin{equation}\label{equ:originalAdditionalConstraints}
 C^\alpha=\pi^\alpha\approx 0,
\end{equation}
whose Poisson-brackets with the original constraints (\ref{equ:originalConstraints}), as well as with any $f(p,q)$, and among themselves, vanish strongly by construction. Let us now apply a point transformation
\begin{equation}\label{equ:pointTransfromaion}
  T_\phi : q_i\to Q_i(q,\phi)
\end{equation}
parametrized by the auxiliary degrees of freedom $\phi_\alpha$, such that $Q_i(q,0)=q_i$, to the original Lagrangian. This transformation is a canonical transformation generated by the generating functional
\begin{equation}\label{equ:generatingFunctional}
 F=Q_i(q,\phi)P^i+\phi_\alpha \Pi^\alpha.
\end{equation}
This is the canonical analogue of the ``Kretschmannization'' procedure in the Lagrangian picture \cite{Kuchar:1990qw}.

Using the shorthand $M^i_j=\frac{\partial Q_j}{\partial q_i}=\frac{\partial \dot Q_j}{\partial \dot q_i}$ as well as $R^\alpha_j:=\frac{\partial Q_j}{\partial \phi_\alpha}=\frac{\partial \dot Q_j}{\partial \dot\phi_\alpha}$ we can denote the canonical transformation from $(q_i,p^i,\phi_\alpha,\pi^\alpha)$ to $(Q_i,P^i,\Phi_\alpha,\Pi^\alpha)$ generated by (\ref{equ:generatingFunctional}) in the compact form\footnote{Alternatively, one can obtain these formulae from the lagrangian, eg.: $p^i=\frac{\partial L}{\partial \dot q_i}=\frac{\partial L}{\partial \dot Q_j}\frac{\partial Q_j}{\partial q_i}=P^jM_j^i $, as $Q$ is the only variable dependent on $q$. In the same way $\pi^\beta=\frac{\partial L}{\partial \dot \Phi_\alpha}\frac{\partial\Phi_\alpha}{\partial \phi_\beta}+\frac{\partial L}{\partial \dot Q_j}\frac{\partial Q_j}{\partial \phi_\beta}  $.}
\begin{equation}\label{equ:canonicalTransformation}
  \begin{array}{rcccl}
   q_i&\to&Q_i&=&Q_i(q,\phi)\\
   p^i&\to&P^i&=&\left(M^{-1}\right)^i_j p^j\\
   \phi_\alpha&\to&\Phi_\alpha&=&\phi_\alpha\\
   \pi^\alpha&\to&\Pi^\alpha&=&\pi^\alpha-R^\alpha_j \left(M^{-1}\right)^j_k p^k.
  \end{array}
\end{equation}
Let us now consider the system of canonically transformed constraints (\ref{equ:originalConstraints}) and (\ref{equ:originalAdditionalConstraints}):
\begin{equation}\label{equ:transformedConstraints}
 \begin{array}{rcl}
   \chi_\rho(p,q) &\to& \chi_\rho\left(Q(q,\phi),P(p,q,\phi)\right)\approx 0\\
   C^\alpha &\to& \pi^\alpha-R^\alpha_j \left(M^{-1}\right)^j_k p^k\approx 0,
  \end{array}
\end{equation}
which is (since obtained using a canonical transformation) still first class. Notice that the previously (almost) trivial constraints $C^\alpha$ now take a quite nontrivial form. To construct a special linking theory, we now assume that we can split the constraints $\chi_\rho\left(Q(q,\phi),P(p,q,\phi)\right)$ into two sets $\chi^1_\alpha(q,p,\phi)$ and $\chi^2_\mu(q,p,\phi)$, where the first set can be solved for $\phi_\alpha$ and the second (weakly) Poisson commutes with $\pi^\alpha$. This happens only in special situations as we will see below. So we can write the constraints (\ref{equ:transformedConstraints}) equivalently as
\begin{equation}\label{equ:constraintsInLinkingForm}
 \begin{array}{rcl}
   0&\approx& \phi_\alpha -\phi^o_\alpha(q,p)\\
   0&\approx& \chi^2_\mu(q,p,\phi)\\
   0&\approx& \pi^\alpha-R^\alpha_j \left(M^{-1}\right)^j_k p^k,
 \end{array}
\end{equation}
which is of the form needed for a special linking theory. We can thus impose the two sets of gauge fixing conditions
\begin{equation}\label{equ:gaugeFixingConditions}
   \phi_\alpha=0 \,\,\textrm{ and }\,\,\pi^\alpha=0,
\end{equation}
which gauge-fix the first and (respectively) last line of (\ref{equ:constraintsInLinkingForm}).

The gauge fixing conditions $\phi_\alpha=0$ can be worked out easily: It only gauge-fixes the constraints $C^\alpha$, whose Lagrange-multiplier  is constrained to vanish, which makes the constraint system independent of $\pi^\alpha$. One can thus perform the phase space reduction by setting $(\phi_\alpha,\pi^\alpha)=(0,\pi_o^\alpha)$ for an arbitrary $\pi_o^\alpha$. Phase space reduction thus reduces to the original gauge theory.

Let us now examine the gauge fixing conditions $\pi^\alpha=0$: These clearly do not (weakly) Poisson-commute with the constraints $\chi^1_\alpha$, because these can be written as $\phi_\alpha -\phi^o_\alpha(q,p)$. To prove that $\pi^\alpha=0$ does not gauge fix any further constraint(s) we assume that there is a subset $\sigma_\rho$ of the constraints $\chi^2_\mu$ and $C^\alpha$ that is gauge-fixed, which implies that the matrix
\begin{equation}
 \left(
   \begin{array}{ccc}
     \{\chi^1,\chi^1\}&\{\chi^1,\pi\}&\{\chi^1,\sigma\}\\
     \{\pi,\chi^1\}&\{\pi,\pi\}&\{\pi,\sigma\}\\
     \{\sigma,\chi^1\}&\{\sigma,\pi\}&\{\sigma,\sigma\}
   \end{array}
 \right) \approx
 \left(
   \begin{array}{ccc}
     0&A&0\\
     -A&0&b\\
     0&-b&0
   \end{array}
 \right)
\end{equation}
is invertible. The block containing $A$ is invertible by assumption and hence the determinant vanishes, since the conjugate block vanishes identically, in contradiction with the assumption. It follows that $\pi^\alpha=0$ can just gauge fix the $\chi^1_\alpha$. This was to be expected of course since the Poisson bracket between the gauge fixing and the constraint of the form $\phi-f(g,\pi)$ is the identity, and thus invertible. Thus, to perform the phase space reduction we trivialize the constraints $\chi^1_\alpha$ and set $(\phi_\alpha,\pi^\alpha)=(\phi_\alpha^o,0)$. This results on the reduced phase space to give the first class constraints
\begin{equation}\label{equ:tradedConstraints}
 \begin{array}{rcl}
   0&\approx&\chi^2_\mu\left(q,p,\phi^o(q,p)\right)\\
   0&\approx&\left(R^\alpha_j \left(M^{-1}\right)^j_k\right)\left(q,p,\phi^o(q,p)\right) p^k=:D^\alpha,
 \end{array}
\end{equation}
and thus effectively traded the constraints $\chi_\alpha^1$ for $D^\alpha$.

In summary we have shown the following proposition
\begin{proposition}\label{prop:constructionPrinciple}
 Given a dynamical system with first class constraints $\chi_\mu$ and a point transformation $q_i\to Q_i(q,\phi)$ parametrized by auxiliary degrees of freedom $\phi_\alpha$ such that a subset $\chi^1_\alpha$ of the constraints can be solved for $\phi_\alpha$ as a function of $(q,p)$ after applying the canonical transformation that implements the point transformation, then the above construction provides a linking theory that provides equivalence with a theory where the $\chi^1_\alpha$ are replaced by constraints $D^\alpha$ as defined in equation (\ref{equ:tradedConstraints}).
\end{proposition}

The phase space reduction of a linking theory can be reversely viewed as an embedding of the equivalent gauge theories into the linking theory. In this picture one has two embeddings $i_{orig.}$ and $i_{dual}$ that embed the original resp. dual gauge theory in the linking theory by
\begin{equation}
 \begin{array}{rcccl}
   i_{orig.}&:&(q,p)&\mapsto&(q,p,0,\pi)\\
   i_{dual}&:&(q,p)&\mapsto&(q,p,\phi^o,0),
 \end{array}
\end{equation}
where $\pi$ denotes an arbitrary constant with components $\pi^\alpha$.

The construction principle described above coincides with Lagrangian best matching in many explicit examples; we find it therefore appropriate to call this construction ``canonical best matching''. In the language of best matching one would call the constraints $C^\alpha\approx 0$ ``Kretschmannization constraints'' and the gauge fixing condition $\pi^\alpha=0$ ``best matching'' or ``free endpoint'' condition. 

\subsection{Linking Theories are not Generic}
 
The canonical best matching procedure described in the previous subsection is a procedure to implement a new symmetry into an original system. However this does not generically lead to a linking theory that can be used to trade symmetries. This can be understood by considering the propagation of the best-matching condition $\pi^\alpha=0$. The Dirac procedure leads to three types of cases for constraint propagation; the generic case is a mixture of the following prototypical cases:
\begin{enumerate}
 \item {\it $\pi^\alpha$ is first class:} The system had the symmetry before best matching. There are two possibilities for this to happen and the generic case is again a mixture of the two:
   \begin{enumerate}
     \item {\it $\pi^\alpha$ is completely irreducible:} This means that the original system is invariant under the new symmetry, so they propagate the new constraints, but the new constraints are not implied by the constraints of the original system. As an example let us consider a particle in 2 dimensions with energy conservation constraint $\chi_o=H(\vec p^2,\vec q^2)-E$, which is by construction invariant under rotations. Let us now best match with respect to rotations around the $3$-axis by extending phase space with an angle $\phi$ and its canonically conjugate momentum $\pi$ using the generating functional $F=P^T.R_3(\phi).q+\phi\Pi$. This results in the Kretchmannization constraint $\chi_3=\pi-T{\epsilon_{3i}}^j q^ip_j$, where $T$ denotes the canonical transform generated by $F$. After applying the canonical transformation, the transformed (or best matched) energy conservation constraint $T\chi_0$ coincides with the original $\chi_o$ and is in particular independent of $\phi$. Thus, the best matching condition $\pi=0$ is automatically propagated. However, the system has changed, because it contains an additional initial value constraint.
     \item {\it $\pi^\alpha$ is completely reducible:} This means that the original system is both invariant under the new symmetry and the initial value constraints of the original system imply the best matching condition. Let us reuse the above example, but now we are adding the rotation constraint $\chi_3$ to the constraints of the initial system, so the Kretschmannization constraint together with the rotation constraint imply the best matching condition $\pi=0$. In this case best matching does not imply a new initial value constraint, so the system has not changed.
   \end{enumerate}
 \item {\it $\pi^\alpha$ generates secondary constraints:} One has to perform the Dirac procedure and ensure propagation of secondary constraints, which can lead to all the cases mentioned so far. Since this case does not differ from the standard Dirac procedure, we refer to textbooks and only consider two simple examples rather than all generality: Consider a particle in three dimensions and a rotationally symmetric energy conservation constraint $\chi_o=H(\vec q^2,\vec p^2)-E\approx 0$ and a single rotation constraint (around the 1-axis) $\chi_1={\epsilon_{1i}}^jq^ip_j$. Let us best-match with respect to rotations around the 2-axis by introducing the auxiliary angle $\phi$ and its canonically conjugate momentum $\pi$. Using the generating functional $F=P^T.R_2(\phi).q+\phi\Pi$ we find the Kretschmannization constraint $\chi_2=\pi-T{\epsilon_{3i}}^j q^ip_j$ and the best matched rotation constraint $\chi_1=T{\epsilon_{1i}}^j q^ip_j$, where $T$ denotes the canonical transform generated by $F$. The best matched energy conservation constraint coincides with $\chi_o$. Now the propagation of the best matching constraint $\pi\approx 0$ implies also $T(\chi_3)\approx 0$, where $T$ denotes the canonical transform generated by $F$, as a secondary (additional) constraint. 

 An example that does not involve rotation symmetry is e.g. a free particle $\chi_o=\vec p^2-E$ that is best matched w.r.t. scale transformations. This is consistent only if $E=0$. However $E=0$ implies the irregular constraint $\chi_o=\vec p^2$, which constrains all momenta.  

 \item {\it $\pi^\alpha$ is second class:} In this case the Dirac procedure stops, as there are no further constraints. One proceeds by solving for the second class constraints explicitly. This is the only case that leads to a linking theory and symmetry trading as e.g. the trading of refoliation invariance for local conformal invariance. A \emph{necessary and sufficient} condition for the existence of a linking theory is that best-matching constraints are first-class amongst themselves (and possibly also with respect to other constraints that one does not want to disturb, e.g. 3-diffeomorphisms) and second class solely with respect to the constraint that one wants to trade for.  
\end{enumerate}
The above examples show that the construction of a linking theory is non-generic. The condition that (a subset of) the best matched original constraints can be uniquely solved for the auxiliary variables $\phi_\alpha$ is indispensable for the existence of a linking theory.

\section{Application to Asymptotically Flat General Relativity}\label{sec:applyToGR}

Let us now apply proposition \ref{prop:equivalence} and the construction leading to proposition \ref{prop:constructionPrinciple} to General Relativity to extend the results of \cite{Gomes:2010fh} to asymptotically flat Cauchy surfaces.

\subsection{Asymptotically Flat Linking Theory}

 To construct the linking gauge theory on a Cauchy-surface $\Sigma=\mathbb R^3$, we must first properly define the appropriate setting. We fix a Euclidean global chart (with radial coordinate $r$) and impose asymptotically flat boundary conditions. We implement this through the fall-off conditions\footnote{These rigid boundary conditions implement asymptotic flatness, but are not the most general boundary conditions for asymptotically flat spacetimes.} of the 3-metric $g_{ab}$ its conjugate momentum density $\pi^{ab}$, the lapse $N$ and shift $N^a$ for the limit $r\to \infty$
\begin{equation}\label{equ:boundary-cond}
 \begin{array}{rclcrcl}
  g_{ab}&\to&\delta_{ab}+\mathcal O(r^{-1}),&&\pi^{ab}&\to&\mathcal O(r^{-2}),\\
  N &\to&c+\mathcal O(r^{-1}),&&N^a&\to&\mathcal O(r^{-1}).
 \end{array}
\end{equation}
We call $\mathcal{C}$ the space of functions on $\Sigma$ with the fall-off rate ascribed to $N$.

We start with the equivalent of \eqref{equ:originalConstraints} and denote the usual ADM constraints as
\begin{equation}
 \begin{array}{rcl}
   S(N)&=&\int d^3x N\left(\frac{\pi^{ab}\pi_{ab}-\frac{1}{2}\pi^2}{\sqrt g}-\sqrt g R\right)\\
   H(v)&=&\int d^3x \pi^{ab}({\mathcal L}_v g)_{ab}.
  \end{array}
  \end{equation}

As before, we now embed the original system into an extended phase space including the auxiliary variables $(\phi,\pi_\phi)$ which we will use for Kretschmannization w.r.t. local conformal transformations. Due to the boundary conditions we assume the scalar $\phi$ falls off as
\begin{equation}
 e^{4\phi}\to 1+\mathcal O(r^{-1})
\end{equation}
for $r\to\infty$ as well as a conjugate momentum density $\pi_\phi$ falling off sufficiently fast at $r\to\infty$. The nontrivial canonical Poisson brackets are
\begin{equation}
  \begin{array}{rcl}
    \{g_{ab}(x),\pi^{cd}(y)\}&=&\delta^{(cd)}_{ab}\delta(x,y)\\
    \{\phi(x),\pi_\phi(y)\}&=&\delta(x,y).
  \end{array}
\end{equation}
The extended phase space for these fields is now given by:
$$(g_{ij},\pi^{ij}, \phi,\pi_\phi)\in\Gamma_{\mbox{\tiny {Ex}}}:=\Gamma_{\mbox{\tiny {Grav}}} \times \Gamma_{{\mbox{\tiny {Conf}}}},$$
where the subscript Grav and Conf denote the original ADM and the conformal extension;
the additional first class constraint to analogous to \eqref{equ:originalAdditionalConstraints} that kills the extension $\Gamma_{{\mbox{\tiny {Conf}}}}$ is:
\begin{equation}
  \pi_\phi\approx 0.
\end{equation}

Following \eqref{equ:generatingFunctional}, we construct the generating function \begin{equation}\label{equ:generatingFunctional2}F_\phi:=\int_\Sigma d^3x \left(g_{ab}(x)e^{4\phi(x)}\Pi^{ab}(x)+\phi(x)\Pi_\phi\right),\end{equation} where capitals denote the transformed variables. We find the canonical transformation analogous to \eqref{equ:canonicalTransformation}:
\begin{equation}
 \begin{array}{rcl}
   g_{ab}(x)&\to& T_\phi g_{ab}(x):=e^{4\phi(x)}g_{ab}(x)\\
   \pi^{ab}(x)&\to&T_\phi \pi^{ab}(x):=e^{-4\phi(x)}\pi^{ab}(x)\\
      \phi(x)&\to&T_\phi \phi(x):=\phi(x)\\
   \pi_\phi(x)&\to&T_\phi \pi_\phi(x):=\pi_\phi(x)-4\pi(x)
 \end{array}
\end{equation}
and subsequently use these transformed variables to construct three sets of constraints: the transformed scalar- and diffeomorphism-constraint of GR  as well as the transform of $\pi_\phi$,
\begin{equation}
 \begin{array}{rcl}
  T_\phi S&=&T_\phi(\frac{\pi^{ab}\pi_{ab}-\frac{1}{2}\pi^2}{\sqrt g}-\sqrt g R)\\
  T_\phi H^a&=&T_\phi(\nabla_b\pi^{ab})\\
  \mathcal{Q}&=&\pi_\phi-4\pi
 \end{array}
\end{equation}
where we have used the shorthand $\pi^{ab}g_{ab}=\pi$. The $\mathcal{Q}$ constraint restricts the functions in $\Gamma_{\mbox{\tiny {Ex}}}$ to be in a one to one relation with the functions on $\Gamma_{\mbox{\tiny Grav}}$. {In other words 
$$0=\{f(g_{ab},\pi^{cd}),\pi_\phi(x)\}=T_\phi\{f(g_{ab},\pi^{cd}),\pi_\phi(x)\}=\{T_\phi f(g_{ab},\pi^{cd}),\pi_\phi(x)-4\pi\}$$ 
is valid and thus $\mathcal{Q}$ holds on the image of $T_\phi$ as applied to functions dependent solely on the original phase space coordinates. It also follows that all $\mathcal{Q}$ weakly commute with the other constraints, and from the fact that $T_\phi$ is a canonical transformation, that all our constraints are first class. The necessary condition that the constraints $T_\phi S$ can be uniquely solved for $\phi$ is a highly nontrivial; we call it privately the ``York miracle'', because it is also the reason why York's procedure works\footnote{Note however that our procedure is not only a different application of the conformal transformations considered by York, but also technically inequivalent to York's procedure.}.

Imposing $\mathcal{Q}$ on the image of $T_\phi$, we can then straightforwardly see that if we smear $T_\phi H^a$:
\be\label{diff}T_\phi H^a(\xi_a):=\int d^3x T_\phi \xi_aH^a \approx \int d^3 x(\pi^{ab}\mathcal{L}_\xi g_{ab}+\pi_\phi \mathcal{L}_\xi\phi)
\ee and thus $T_\phi H^a$ still generates diffeomorphisms  in the extended space. The notation used here is such that the specific smearing of functions is denoted by $F(f)$, whereas more general functionals are denoted by $F[f]$.
Since we regard the linking theory as the central object in this paper, we could of course have started from a canonical Lagrangian which had the constraint in the (smeared) form of the R.H.S. of \eqref{diff} rather than the transformed GR diffeomorphism constraint. However in his case one would have to check whether the constraints are first class, which can be confirmed by a simple direct calculation.

Using a scalar Lagrange-multiplier $\rho$, which is supposed to fall off as $\mathcal O(r^{-1})$ as $r\to\infty$, we define the total Hamiltonian\footnote{We should for general purposes add a regularizing boundary term
to the total Hamiltonian, as it diverges in the present form. However since this does not impinge on either the equations of motion nor on the constraints, we omit it in order to avoid cluttering  the paper.}
\be\label{Hamiltonian} H_{\mbox{\tiny {Total}}}=\int d^3x [N(x)T_\phi S(x)+\xi^a(x)T_\phi H_a(x)+\rho(x)\mathcal{Q}(x)]
\ee
We do not explicitly topologize phase space for now and only later assume that we can turn it into a Banach space compatible with the Poisson bracket.

This completely defines the linking $T_L$ as contained in the previous section. We define the linking theory as the gauge theory defined in this section together with the two sets of gauge-fixing conditions and constraint sets
\begin{eqnarray}
\mbox{Constraints}&:& \mathcal X_1={Q}\,\,\,\textrm{ and }\,\,\, \mathcal X_2=\phi-\phi_o\,\,\,\textrm{ and }\,\,\, \mathcal X_0=T_\phi H^a\cup \langle N_0 T_{\phi}S\rangle\nonumber\\
\mbox{Gauge fixing}&:&\Sigma_1=\{\pi_\phi(x)=0\}_{x\in\Sigma} \,\,\,\textrm{ and }\,\,\, \Sigma_2=\{\phi(x)=0\}_{x\in\Sigma}\label{gauge conditions},\end{eqnarray}
 where $\phi_0$ and $N_0$ will be specified shortly in a way that ensures that $\phi-\phi_o$ combined with $\langle N_0 T_{\phi_o}S\rangle$ is equivalent to $T_\phi S(x)$ at the surface $\pi_\phi\equiv0$ for the boundary conditions given in (\ref{equ:boundary-cond}).

\subsection{Recovering General Relativity}

The only nonvanishing Poisson-bracket of the gauge fixing condition $\phi(x)=0$ with the constraints of the linking theory is
\begin{equation}
 \{\phi(x),Q(\rho)\}=\rho(x),
\end{equation}
which determines the Lagrange-multiplier $\rho(x)=0$, which eliminates $\pi_\phi$ from the theory. We can thus perform the phase space reduction by setting $\phi(x)=0,\pi_\phi(x)=0$. The constraint $Q(0)$ is empty. Moreover, for phase space functions independent of $\phi,\pi_\phi$ one finds that the Dirac-bracket coincides with the canonical Poisson bracket and the constraints on the reduced phase space are
\begin{equation}
S(x)~~\mbox{and}~~ H^a(x)
\end{equation}
The resulting gauge theory is thus ADM gravity. In this case, we can follow through from \eqref{gauge conditions}, arriving at, in the language of proposition \ref{prop:equivalence}, $\rho \equiv 0$ and $f_i\approx 0$ equivalent to $S(x)\approx 0$. We have lost the freedom to fix $\rho$, but retained the freedom to fix the lapse.

\subsection{Recovering Shape Dynamics}

Our main aim in this section will be to prove that part of the scalar constraints can be written in the form $\phi-\phi_0(g,\pi)\approx 0$ at the gauge-fixing surface $\pi_\phi\equiv 0$, and then use the results of section 2.

The only weakly non-vanishing Poisson-bracket of the gauge-fixing condition $\pi_\phi(x)=0$ with the constraints of the linking theory is $\{T_\phi S(N),\pi_\phi(x)\}=4T_\phi\{S(N),\pi(x)\}$, which leads to
\begin{equation}
 \{ S(N),\pi(x)\}=
 2(\nabla^2N-NR)\sqrt g-\frac 3 2 NS\approx 2\sqrt g(\nabla^2-R)N
\end{equation}
The differential operator
\begin{equation}
\label{Delta}\Delta=\nabla^2-R
\end{equation}
for the  given boundary conditions, is an invertible operator.  So, for each value of $c$ in  (\ref{equ:boundary-cond}), we have the unique kernel \be\label{N_0}N^c_0[g,\pi]\neq 0\ee
Furthermore, from the linearity of the equation we have $N^c_0=cN^1_0$.

Since $\mathcal{T}_{\phi'}\{ S(N),\pi(x)\}=\{\mathcal{T}_{\phi}S(N),\pi_\phi\}_{|\phi=\phi'}$,  we can extend this result to all of extended phase space, and thus we have that indeed there is a one-dimensional space of admissible smearings, whose contraction with $\mathcal{T}S(x)$ remains first class with respect to all the other constraints. We denote a particular generator of this constraint by
\be\label{H_gf}
H_{\mbox{\tiny g.f.}}:=T_\phi S(N_0).
\ee
where we use the notation that whenever the superscript is omitted it is being set to 1.

We use the  freedom in $c$ present in the boundary conditions to select out a unique generator for our first class constraints. This will also be important in the next section where we show that the remaining scalar constraints can be put into the form  $\phi-\phi_0(\Gamma_{\mbox{\tiny Grav}})$.  
 We note that a regularizing term  must appear here as in almost every other procedure involving asymptotic spaces in GR.\footnote{E.g. one must add a  ``boundary term'' consisting of the extrinsic curvature of a sphere embedded in Minkowski space to the usual  GR Hamiltonian. } In this sense, we have:
\be\label{equ:normalization}\lim_{R\rightarrow\infty}\frac{1}{V_R}\int_{r<R} N^c_0\sqrt {|g|}d^3 x=1 ~~ \Leftrightarrow c=1
\ee
where the integral is taken over expanding balls of radius $r$ embedded in Euclidean space. This integral can be computed through simple estimates. Thus we have that 
\be \mean{N_0}=1
\ee
by the limiting procedure defined above.

Now, we do not fix the lapse gauge to be given by $N_0$, but we separate the constraints into a first class part, given by
$$
\mbox{\bf First class:}~~~\{~~T_\phi S(N_0), \{\mathcal{Q}(x),x\in\Sigma\}, \{T_\phi H^a(x),x\in\Sigma\}~~\}
$$
 and a purely(as we will see shortly)  second class part, given by
$$
  \mbox{\bf Second class:}~~~~\{~~\{\widetilde{T_\phi S}(x):=\mathcal{T}_\phi S(x)-\mathcal{T}_\phi S(N_0)\frac{\sqrt{|g|}}{V},x\in\Sigma\}, \{\pi_\phi(x),x\in\Sigma\}~~\}.
$$
where we define:
\be \frac{\mathcal{T}_\phi S(N_0)}{V}:=\lim_{R\rightarrow\infty}\frac{1}{V_R}\int_{r<R} \mathcal{T}_\phi S(x) N_0(x)\sqrt {|g|}d^3 x
\ee

We will discuss the affirmation that $\widetilde{T_\phi S}$ is indeed purely second class in the next section.

\subsubsection{Constraint Surface for Shape dynamics}

Now we  show that in the asymptotically flat case the constraint $\widetilde{{ T}_\phi S}$ is equivalent to a constraint of the form $\phi-\phi_0(\Gamma_{\mbox{\tiny Grav}})$.

 We have that
 \begin{equation}{ \mathcal{T}_\phi S}(x):\Gamma\times T^*(\mathcal{C}_r)\rightarrow C^\infty(M),\end{equation}  Since these equations do not depend on $\pi_\phi$, we can fix $\pi_\phi(x)=f(x)$. Then
  \begin{equation} { {{ T}_\phi}S}(x)_{\pi_\phi=f(x)}:\Gamma\times \mathcal{C}_r\rightarrow C^\infty(M).
\end{equation}

Consider the linear operator:
   $$\delta_{\mathcal{C}}{\mathcal {T}_\phi}S(g_0,\pi_0,\pi_\phi^0)_{|\phi=0}: {T}_1^*(\mathcal{C}_r)\rightarrow C^\infty(M).
   $$ where $T^*_xN$ denotes the cotangent space at $x\in N$, and, as in usual partial derivatives, one holds the coordinates $(g,\pi,\pi_\phi)$ fixed. As usual, we can identify $ {T}_1^*(\mathcal{C}_r)$ with $ \mathcal{C}_r$ itself.  
We will omit from now on the ``initial" point $(g_0,\pi_0,\pi_\phi^0)$ where we take the derivative.
Thus
   $$\delta_{\mathcal{C}}{{ T}_\phi}S_{|\phi=0}:\mathcal{C}_r\rightarrow C^\infty(M).
   $$
    Clearly $\widetilde { T}_\phi S(N_0)=0$, which means indeed it lives in the dual space of the quotient of $C^\infty(M)$ by $N_0$.  Effectively, we must subtract from any $N\in\mathcal{C}$ the function $ N_0$ given by \eqref{N_0}.

   Consider the linear self adjoint elliptic operator we presently have:
\begin{equation}\label{iso} \delta_{\mathcal{C}}{{ T}_\phi}S_{|\phi=0}:=\frac{\delta {{{T}_\phi}\mathcal{H}(x)}}{\delta\phi(y)}_{|\phi=0}=\{{{ T}_\phi}H(x),\pi_\phi(y)\}_{|\phi=0}=\Delta(x)\delta(x,y)
\end{equation}
Although in this case the difference is immaterial (as we are dealing with a self-adjoint operator), contraction of the above operator on the $y$ index would imply that we are using the adjoint (see \cite{Fisher_Marsden:Einstein_Centenary}): 
$$  (\delta_{\mathcal{C}} \mathcal {T}_\phi S)^*\cdot N= \Delta N
$$ where $\Delta$ is given by \eqref{Delta}.

If we then use  the inner product in the $C^\infty(M)$:
\be \label{equ:innerprod} \langle f,m\rangle_{C^\infty(M)}:=\int d^3x\sqrt g fm\ee
we have
   \begin{equation}\{\mathcal {T}_\phi S(N),\pi_\phi(\rho)\}\approx\langle (\delta_{\mathcal{C}} \mathcal {T}_\phi S)\cdot\rho,N\rangle=\langle (\delta_{\mathcal{C}} \mathcal {T}_\phi S)^*\cdot N,\rho\rangle
 \end{equation}
But note that although there exists a unique solution for $\Delta N=0$ \emph{if} $N\rightarrow 1$, the domain of the operator  $\mathcal {T}_\phi S)^*$ is $C^\infty(M)$. So we do not know a priori whether the entire kernel is indeed generated by $N_0$.  To overcome this we simply use  the Fredholm alternative and self-adjointness\footnote{In fact,  the Fredholm alternative has subtleties when in the non-compact manifold case. To overcome this, one should instead use Christodoulou's `method of continuity', which also uses a limiting procedure akin to the one we have used above \eqref{equ:normalization}.}
 \be\label{equ:splitting2}C^\infty(M)\simeq  \text{Im}(\delta_{\mathcal{C}} \mathcal {T}_\phi S) \oplus \text{Ker}(\delta_{\mathcal{C}} \mathcal {T}_\phi S)^*= \text{Im}(\delta_{\mathcal{C}} \mathcal {T}_\phi S) \oplus \text{Ker}(\delta_{\mathcal{C}} \mathcal {T}_\phi S)
 \ee
But we know there exists a unique solution to $\Delta\rho=0$ given the boundary conditions imposed on the space $\mathcal{C}_r$, which is the domain of $\rho$.  Thus we have a unique solution $\rho_0$, which therefore must be given by $\rho_0=N_0$,  as it has the same boundary conditions as the space of smearings. The splitting is then given by
\be \text{Im}(\delta_{\mathcal{C}} \mathcal {T}_\phi S) \oplus \text{Ker}(\delta_{\mathcal{C}} \mathcal {T}_\phi S)=\text{Im}(\delta_{\mathcal{C}} \mathcal {T}_\phi S) \oplus N_0
\ee
Using the self-adjointness, we bypassed complicated issues regarding the relations between the range of  $\mathcal {T}_\phi S$, given by $C^\infty(M)$, and the space of admissible lapse functions. To wit, in the closed compact case no restriction on the space of admissible $N$ was necessary, and we could directly derive the appropriate splitting. Here, we would have no straightforward way to know that indeed the whole of $\text{Ker}(\delta_{\mathcal{C}} \mathcal {T}_\phi S)^*\subset C^\infty(M)$ was generated by $N_0$, only that its linear intersection with the space of admissible lapses would be given by $N_0$. However as $\text{Ker}(\delta_{\mathcal{C}} \mathcal {T}_\phi S)^*=\text{Ker}(\delta_{\mathcal{C}} \mathcal {T}_\phi S)$, and we know $\delta_{\mathcal{C}} \mathcal {T}_\phi S$ acts on the space $\mathcal{C}_r$, we have the unique solution $\rho_0=N_0$ and can perform the required splitting.

 By the canonical transformation properties of $\mathcal{T}_\phi$, one can extend this construction to arbitrary $\phi$. We have thus proven
\begin{proposition}\label{prop2}
The linear map given by $\delta_{\mathcal{C}}\widetilde{\mathcal {T}_\phi S}(x):T_1(\mathcal{C}_r)\rightarrow \text{Im}(\delta_{\mathcal{C}} \mathcal {T}_\phi S)$
where $\widetilde{\mathcal {T}_\phi S}(x)=\mathcal {T}_\phi S(x)-H_{\mbox{\tiny gl}}\frac{\sqrt{g}}{V}:\mathcal{C}_r\times\Gamma\rightarrow \text{Im}(\delta_{\mathcal{C}} \mathcal {T}_\phi S) $, is a toplinear isomorphism for all $(\phi, g,\pi)$.
  \end{proposition}
  We have shown that it is a linear continuous bijection, and hence a topological linear isomorphism \cite{Lang}. $\square$.

 It is important that there is a non-zero homogeneous solution to the $\Delta$ operator, so we are left with a first class component of ${ T}_\phi S$. This is the reason for using not full conformal transformations in the compact case, but only those that preserve the total spatial volume. The analogous restriction arises from the fall-off conditions in the present case.
The construction of the theory on the constraint surface and the further fixing of the gauge therefore proceeds in the same manner as was shown in the constant mean curvature case. Of course  instead of having constant mean curvature slicing, we now have the maximal slicing $\pi=0$.

   Thus not only can we form the Dirac bracket using $\{\widetilde{T_\phi H}(x),\pi_\phi(y)\}^{-1}$,  but we can now use the implicit function theorem for Banach spaces to assert that
\begin{proposition}\label{prop}
 There exists a unique $\phi_0:\Gamma\rightarrow \mathcal{C}$ such that \end{proposition}
 $$ (\widetilde{\mathcal{T}_\phi S})^{-1}(0)=\{(g_{ij},\pi^{ij},
 \phi_0[g_{ij},\pi^{ij}],\pi_\phi)~|~(g_{ij},\pi^{ij})\in\Gamma_{\mbox{\tiny {Grav}}}\}.
$$

As a last comment, we also note that unlike the compact case, we now have that the global leftover Hamiltonian is given by:
\be\label{equ:global_Hamiltonian}H_{\mbox{\tiny gl}}:= \mathcal{T}_{\phi_0}S(N_0)= V\mathcal{T}_{\phi_0}S(x)
\ee where the regularizing volume appears in the rhs. 

\subsubsection{Construct the theory on the constraint surface}

We now have a surface in $\Gamma_{\mbox{\tiny {Ex}}}$, defined by $\pi_\phi=0$ and $\phi=\phi_0$, on which $\widetilde{T\mathcal{H}}=0$, and whose intrinsic coordinates are $g_{ij},\pi^{ij}$.
Furthermore, the Dirac bracket on the surface exists and,  and we now have the symplectic structure on the constraint surface
\begin{equation}
  \label{reduced Pb}\{\cdot,\cdot\}_{|{\mbox\tiny{reduced}}}:={\{\cdot,\cdot\}^{\Gamma_{\mbox{\tiny {Ex}}}}_{\mbox{\tiny{DB}}}}
    =\{\cdot_{|\phi=\phi_0,\pi_\phi=0},\cdot_{|\phi=\phi_0,\pi_\phi=0}\}.
\end{equation}Equivalently, for phase space functions independent of $\phi, \pi_\phi$, analogously to \eqref{Dirac bracket}:
\begin{multline} {\{F_1(x),F_2(y)\}^{\Gamma_{\mbox{\tiny {Ex}}}}_{\mbox{\tiny{DB}}}}_{|\phi=\phi_0,\pi_\phi=0}=\\
\{F_1(x),F_2(y)\}+\{F_1(x),(\phi-\phi_0)(x')\}\{\pi_\phi(x'),F_2(y)\}-\{F_1(x),\pi_\phi(x')\}\{(\phi_0-\phi)(x'),F_2(y)\}
\\ {=\{F_1(x),F_2(y)\}}\end{multline}
where the repeated variable $x'$ is integrated over.

One can immediately see from \eqref{reduced Pb} that the first class constraints ${4\pi},T\mathcal{H}^a$ and $ \langle T N_0\mathcal{H}\rangle$ remain first class. Alternatively, for the diffeomorphism constraint, one can directly observe from \eqref{diff} that setting $\pi_\phi=0$ reduces $T_\phi H^a\rightarrow H^a$, yielding the usual diffeomorphism constraint.  We thus find the total Shape Dynamics Hamiltonian
\begin{equation}
  H_{\text{dual}}=\mathcal N\langle T_{\phi_0} N_0\mathcal{H}\rangle+\int_\Sigma d^3 x \left(\lambda(x) 4\pi(x)+\rho^a(x)H_a(x)\right)
\end{equation}
in the ADM phase space $\Gamma$ with the first class constraints
\begin{equation}
  \langle T_{\phi_0}N_0\mathcal{H}\rangle,~{4\pi},\mathcal{H}^a.
\end{equation} We have thus effectively fixed the gauge $N=N_0$ at the surface $\phi=\phi_0$. We have lost the freedom to fix the lapse, but retained the freedom to choose the conformal Lagrange multiplier $\rho$.

The non-zero part of the constraint algebra is given by:
\begin{eqnarray}
\{H^a(\eta_a),H^b(\xi_b)\}&=&H^a([\vec\xi,\vec\eta]_a)\nonumber\\
\label{equ:constraintAlgebra}\{\langle T_{\phi_0}N_0\mathcal{H}\rangle,\pi(\rho)\}&=&0\\
\{H^a(\xi_a),\pi(\rho))\}&=&\pi(\mathcal{L}_\xi\rho)\nonumber
\end{eqnarray}

\subsection{Further fixing of the gauge}

As an explicit check to see whether we indeed have the same theory, we can further gauge fix both ADM and Shape Dynamics to a system which possesses exactly the same gauge fixed Hamiltonian.
To do this, we merely input the further gauge fixing $S(x)=0$ in Shape Dynamics and $\pi=0$ in ADM. On the Shape Dynamics side, we have that restriction to the gauge fixing surface implies that we are over $(g,\pi)$ for which  $\phi_0(g,\pi)=0$. On the GR side, we have maximal slicing, which requires that $N=N_0$, and thus we arrive explicitly at GR in maximal CMC gauge form both sides and have thus verified that the trajectories of the two theories are the same.

\subsection{Connection with previous work}\label{sec:connections}

Let us relate the content of the present paper with previous work, in particular \cite{Anderson:2002ey,Barbour:1999mf,Barbou:CS_plus_V,Barbour:shape_dynamics} before we justify the usefulness of  linking theories and the extension of phase space. From the start, let us make it clear that \emph{the present theory is in no way equivalent to any of previous work}. The distinction does not only arise due to a different formalism (Hamiltonian vs Lagrangian), but in the fact that the theories are de facto different. Shape Dynamics, in contrast to previous work, possesses local conformal invariance, and its Hamiltonian \emph{is not} just ADM in CMC gauge. In our notation, the ADM in CMC gauge would be just $S(N_0)$, where $N_0$ is the appropriate solution of a given lapse fixing equation. From \eqref{equ:global_Hamiltonian}, we can see that this would only be the case for SD if $\phi_o=1$. The presence of the functional $\phi_o$ is fundamental for conformal invariance.  

 These previous papers were concerned with the establishment of conformal transformations in General Relativity from truly relational first principles, but the work was not concerned with trading symmetries. This is also the reason for using two different technical approaches, because the relational picture is more transparent in the Lagrangian formulation, which was mostly used in previous work, while symmetry trading is more transparent in the Hamiltonian picture. We thus see our work as an extension of references \cite{Anderson:2002ey,Barbour:1999mf,Barbou:CS_plus_V,Barbour:shape_dynamics} by giving a clean Hamiltonian description and identifying the underlying reason for the surprising results of the previous papers as the symmetry trading between refoliation- and local spatial conformal symmetry. However, without performing a complete Dirac analysis it was not in their possibilities to obtain the symmetry trading and the leftover Hamiltonian. A second extension of previous work is the technical explanation of symmetry trading in terms of linking theories, where it can be understood straightforwardly.  

Let us now consider the usefulness of linking theories in more detail and in particular consider whether it should be considered as "mere formalism". There are two sides to the argument: One necessity and the other is simplicity and practicality. It turns out that linking theories are not conceptually absolutely necessary, in the sense that the linking structure itself  can be explained on the original phase space $\Gamma(p,q)$ once one knows both theories exist, as follows:

Let us again consider three disjoint sets of constraints $\{\chi^1_\alpha(p,q)\}_{\alpha \in \mathcal A},\{\chi_2^\alpha(p,q)\}_{\alpha \in \mathcal A},\{\chi^3_\mu(p,q)\}_{\mu \in \mathcal M}$. We require that the two sets $\{\chi^1_\alpha(p,q)\}_{\alpha \in \mathcal A} \cup \{\chi^3_\mu(p,q)\}_{\mu \in \mathcal M}$ and $\{\chi_2^\alpha(p,q)\}_{\alpha \in \mathcal A} \cup \{\chi^3_\mu(p,q)\}_{\mu \in \mathcal M}$ are each first class by themselves, so there are two distinct gauge theories on $\Gamma(p,q)$. Now we require in addition the linking condition that the Dirac matrix $\{\chi^1_\alpha,\chi_2^\beta\}$ is invertible. This allows us to impose the set $\{\chi_2^\alpha(p,q)\}_{\alpha \in \mathcal A}$ as a partial gauge fixing for the first gauge theory and vice versa to impose the set  $\{\chi^1_\alpha(p,q)\}_{\alpha \in \mathcal A}$ as a partial gauge fixing for the second gauge theory. The two partially gauge fixed theories then have the same initial value problem and the same equations of motion, so they can be identified with one another.

However, working out the reduced phase space, where the two gauge theories coincide is in general not feasible, hence the use of linking theories is definitely more practical. As an example let us consider what one would have to do if we wanted to establish the duality between General Relativity and Shape Dynamics entirely on the ADM phase space $T^*\mbox{Riem}$:

It is possible to determine $\tilde S$ directly in $T^*\mbox{Riem}$. But to find $\hg$ we exploited the conformal transformations, eventually finding a functional $\phi_0$ which solved for $\tilde S$ and thus yielded our global Hamiltonian $\hg$. For this we would have to complete the following steps which are straightforward in the linking theory and appear in round brackets of the form (LT: ...):

 \begin{itemize}
\item Step 1: Impose $D(x)=0$ as a gauge-fixing in ADM (LT: impose $\pi_\phi=0$). 
\item Step 2: Find the variable canonically conjugate to $D(x)$, let us call it $q_D$ (LT: this is just $\phi$). Already at this step the procedure is already practically impossible in most cases, as one cannot find such a canonical conjugate without the introduction of an extra variable.
 \item Step 3:  Find variables that Poisson commute with both $D(x)$ and its canonical conjugate, $q_D$. These would already be vpct invariant variables (LT: these are just $(g_{ab},\pi{ab})$).
\item Step 4: find the purely second class combination of the constraints with respect to $D=0$. This is given by $\tilde S$. (LT: this is given by $\widetilde T_\phi S$).
\item  Step 5:  Finally, show  that indeed we could solve all but one of the original scalar constraints $S(x)$ for $q_D$ as a function of the remaining variables (LT: this solution is given by $\phi^0[g,\pi,x)$).
\end{itemize}

As all the quantities would be expressed in terms of variables that Poisson commuted with $D(x)$ we would automatically have a vpct invariant theory without going into extended phase space. It seems that this is roughly what Dirac had in mind in \cite{Dirac:CMC_fixing}, albeit solely from a gauge fixing point of view, i.e. not involving a ``conformal transformation" conceptual background, and also limiting the analysis to asymptotically flat space. 

The linking theory formalism is thus not physically necessary in principle, but in practice extremely useful, particularly when one considers more complicated models than pure gravity. Let us conclude the case for linking theories with the remark that going to an extended phase space leads to a clearer physical/mathematical  description in many other well-known settings. Usual gauge theories can be described as local field theories precisely because one includes gauge degrees of freedom, while the theory is nonlocal in terms of physical degrees of freedom. Another example is the BRST-extension, which gives a clear picture of gauge invariance in terms of algebraic geometry (cohomology) by introducing ghost degrees of freedom. The analogy of BRST with linking theories goes even further, as both involve an extension of phase space that transforms a system of constraints with an algebra that is only closed on the constraint surface to one that is closed on the entire extended phase space.  We cannot omit the work of Stueckelberg himself, who already in the late 40's  introduced a scalar field into an Abelian gauge theory, making it massive but also preserving gauge invariance.\footnote{It can arguably be said that the mechanism of spontaneous symmetry breaking was in this way hinted at much before the work of Anderson and Higgs.} The so-called ``Stueckelberg mechanism" is usually described as the introduction of new fields extraneous fields in order to reveal a symmetry of a gauge--fixed theory,  on par with what is done for SD and the general case described in the present paper. 

\section{Lagrangian Picture}\label{sec:Lagrangian}

Let us consider the Lagrangian of the linking gauge theory in an attempt to relate the local degrees of Shape Dynamics with those of General Relativity. The local degrees of freedom of standard General Relativity are given by the ADM-decomposition of a 4-metric, i.e. a 3-metric, shift vector field and lapse field, while shape dynamics is a local theory of a 3-metric, a shift vector field, the conformal field $\phi$ and the conformal Lagrange-multiplier $\rho$. While the 3-metric and shift vector field are naturally identified, one needs to consider the Euler-Lagrange-equations to investigate further.

Using $D(\xi)=\int d^3x H^a(x)\xi_a(x)$, $C(\rho)=\int d^3x \mathcal Q(x) \rho(x)$ and the supermetric $G_{abcd}=g_{ac}g_{bd}-\frac 1 2 g_{ab}g_{cd}$ we can write the action for the linking theory in canonical form as
\begin{equation} \label{equ:linking-action}
 \begin{array}{rcl}
  S&=&\int dt\left(d^3x \left(\dot g_{ab}\pi^{ab}+\dot\phi \pi_\phi\right) -\left(T_\phi S[N]+D[\xi]+C[\rho]\right)\right)\\
   &=&\int dtd^3 x\left( \frac{1}{4N}G^{abcd}(\dot g_{ab}-\mathcal{L}_\xi g_{ab}-\rho g_{ab})(\dot g_{cd}-\mathcal{L}_\xi g_{cd}-\rho g_{cd})+N T_\phi R\right),
 \end{array}
\end{equation}
where we used the equations of motion
\begin{equation}
 \begin{array}{rcl}
   \dot g_{ab}&=&2N \pi^{cd}G_{abcd}+\mathcal L_\xi g_{ab}+\rho g_{ab}\\
   \dot \phi&=&-\rho-\mathcal{L}_\xi\phi
 \end{array}
\end{equation}
to eliminate the momenta. Coming purely from the canonical Lagrangian one could now think that one could find an equation that relates the lapse and the conformal Lagrange multiplier, which basically means to relate the local speed of time to local speed of scale. To see that this is not  possible, we consider the  the general construction principle for linking theories as explained in section \ref{sec:constructionPrinciple}.  We start with the Gauss-Codazzi split of the Einstein-Hilbert action
\begin{equation}
 S=\int dt d^3 x\sqrt{|g|}\left(\frac 1{4N}\left(\dot g_{ab}-(\mathcal L_\xi g)_{ab}\right)G^{abcd}\left(\dot g_{cd}-(\mathcal L_\xi g)_{cd}\right)+N R[g]\right)
\end{equation}
and use the transfromation $g_{ab}(x)\to T_\phi g_{ab}(x):= e^{4\phi(x)}g_{ab}(x)$, which again yields the action for the linking theory, i.e. the second line of (\ref{equ:linking-action}), with $-\rho$ replaced by $\dot \phi$ as it should be
\begin{equation}
  S=\int dtd^3 x\left( \frac{1}{4N}G^{abcd}(\dot g_{ab}-\mathcal{L}_\xi g_{ab}+\dot\phi g_{ab})(\dot g_{cd}-\mathcal{L}_\xi g_{cd}+\dot\phi g_{cd})+N T_\phi R\right).
\end{equation}
Note that the charge $Q$ (as it came out of the canonical procedure in the previous section) comes out as a primary constraint from the Legendre transform of this Lagrangian as it was to be expected from section \ref{sec:constructionPrinciple}. However, there is no relationship possible between the Lagrange multiplier $\rho$ of the conformal constraint and the lapse $N$, neither in the linking theory (where both Lagrange multipliers are free) nor Shape Dynamics (where the lapse is fixed but $\rho$ is free). One could now argue that one can write a relationship between $N$ and $\phi, \dot\phi$ by imposing the constraint $\pi_\phi=0$. But this choice, as we showed in the previous section, fixes $\phi=\phi_0, \dot\phi=\dot\phi_0$. We thus find for this case that the Lagrangian for Shape Dynamics is written as the second line of equation (\ref{equ:linking-action}) with $\phi=\phi_0$ and $-\rho=\dot \phi_0$, which admits no dynamical relation anymore.

\section{Conclusion}

This paper is intended to make the rather involved construction presented in \cite{Gomes:2010fh}, which established the equivalence between General Relativity and Shape dynamics in the compact without boundary case, more transparent and at the same time to extend the equivalence of General Relativity and Shape Dynamics to the asymptotically flat case as an example for the general applicability of this procedure. (See the appendix for a brief discussion of the adaptation of the arguments in this paper to the compact without boundary case.)

We started by showing how an equivalence of gauge theories follows from the existence of a linking gauge theory on an extended phase space $\Gamma\times \tilde \Gamma$.  One can sketch the construction of a pair of equivalent gauge theories A and B on a reduced phase space $\Gamma$ as follows
\begin{equation}
  \begin{array}{ccccc}
     &\textrm{partial gauge fixing}&&\textrm{partial gauge fixing}&\\
     \textrm{theory A}&\longleftarrow&\textrm{linking theory}&\longrightarrow&\textrm{theory B}\\
      \textrm{on }\Gamma \times \tilde\Gamma& \phi_I=0&\textrm{ on }\Gamma \times \tilde\Gamma&\pi_\phi^I=0& \textrm{on }\Gamma \times \tilde\Gamma\\
     \downarrow &&&& \downarrow\\
     \textrm{ reduced } &&&& \textrm{ reduced }\\
     \textrm{ theory A} &&&& \textrm{ theory B}\\
     \textrm{on }\Gamma &&&& \textrm{on }\Gamma\\
     &\longrightarrow&\textrm{Dictionary}&\longleftarrow&\\
     &&\textrm{ on }\Gamma_{red},&&
  \end{array}
\end{equation}
where $\phi_I$ and $\pi_\phi^I$ is a canonical pair coordinatizing $\tilde \Gamma$ and the \lq\lq Dictionary\rq\rq is a further gauge fixing of the two equivalent theories such that the two theories coincide. The dictionary can be used to easily identify trajectories of the equivalent theories with one another. 

The dictionary\footnote{We use the word dictionary, because only this gauge allows for a direct identification of GR and SD trajectories and thus translates between the ''spacetime language`` used to interpret Hamiltonian GR and the ''intrinsic language`` used to interpret SD.} between General Relativity and Shape Dynamics is General Relativity in CMC gauge, where by construction the trajectories of the two theories coincide. However, since both theories arise as partial gauge fixings of the linking theory, one does not have to impose the existence of the Dictionary to prove equivalence, which may open the possibility that Shape Dynamics could admit trajectories that cannot be translated into CMC-foliable solutions of General Relativity.

We proceeded by giving a general construction principle for linking theories in section \ref{sec:constructionPrinciple}. This simple construction principle is, although equivalent to the previously applied St\"uckelberg formalism, technically much simpler to handle and serves as the basis for ongoing work on coupling Shape Dynamics to matter, which we intend to report on shortly.

We followed this simple construction principle and found a linking gauge theory that established equivalence of General Relativity and Shape Dynamics in the asymptotically flat case and thus showed their equivalence as gauge theories. This theory, like its counterpart on a compact Cauchy surface without boundary, lives on the same phase space as General Relativity, is invariant under spatial diffeomorphisms and has traded refoliation invariance for spatial conformal invariance and a global Hamiltonian.

Taking Shape Dynamics as a theory in its own right, i.e. without requiring that it is everywhere equivalent to General Relativity, we have no reason to impose positive definiteness of the lapse $N_0$ anymore, since the forward-evolution in time is given by the positivity of the Lagrange-multiplier of the global Hamiltonian of Shape Dynamics. We can thus merely require that a lapse $N_0$ exists as a not identically vanishing function\footnote{This simplifies the existence discussion compared to General Relativity and also makes proof of the existence of the $\phi_0$ functional extremely easy, a simple application of the implicit function theorem, and goes in contrast to the more elaborate original constructions of solutions to the Lichnerowicz-York equation and the lapse fixing equation for CMC.}. As of now, we do not have a classical example where the conformal gauge freedom shows a new effect. This is however different in the quantum theory, where metric Shape Dynamics can be Dirac quantized on the 2+1 dimensional torus universe \cite{Budd:2011er}, while the metric Wheler-DeWitt equation is not well defined.

\subsection*{Acknowledgements}

We wish to thank Julian Barbour and Sean Gryb for helpful discussions. HG thanks the Perimeter Institute for hospitality. Research at the Perimeter Institute is supported in part by the Government of Canada through NSERC and by the Province of Ontario through MEDT. HG was supported in part by the U.S.
Department of Energy under grant DE-FG02-91ER40674.

\section*{Appendix: Differences from the compact $\Sigma$ case}

In this paper we considered asymptotically flat boundary conditions to demonstrate the generality of our procedure. However, imposing boundary conditions is definitely in conflict with a relational picture. We thus want to supplement the paper with a discussion of the truely relational compact without boundary case. To this end we briefly discuss the main differences between the asymptotically flat boundary condition and compact Cauchy surfaces without boundary.
\begin{enumerate}
\item As mentioned in the text, we must use not full conformal transformations, but volume preserving conformal ones. There exists a simple surjective map that projects $\phi\mapsto \hat\phi$. The generating functional equivalent to \eqref{equ:generatingFunctional} is then given exactly by the same form, but with $\hat\phi$. This brings in a mild complication to the transformation properties of the momentum, which now goes as:
     $$T_{\hat\phi}\pi^{ab}=e^{-4\hat\phi}(\pi^{ab}-\frac{g^{ab}}{3}\sqrt {g}\langle \pi\rangle (1-e^{6\hat\phi}))
     .$$This complication is traded with that of giving boundary conditions of the asymptotically flat case. It is necessary in order to have some combination of the linking theory scalar constraint $\{ T_{\hat\phi}S(x), x\in\Sigma\}$ that is not fixed by (remains first class wrt) the condition $\pi_\phi=0$. This allows shape dynamics to have a true Hamiltonian, and be matched to ADM in something other than the frozen lapse regime.
\item By using such conformal transformations, one finds that we still have a canonical transformation, but the charge which identifies the linking theory in extended phase space to ADM, is given by $\mathcal{Q}=\pi_\phi-\mathcal{D}$ where $\mathcal{D}:=4(\pi-\langle\pi\rangle\sqrt g)$. It can be seen by explicit calculation to commute with any $T_{\hat\phi} f$, for $f$ a function of $g,\pi$.
    \item The diffeomorphism constraint in the linking theory $T_\phi H^a(\xi_a)$, although more complicated,  can still be explicitly calculated as follows (in smeared density form):
        \begin{eqnarray*}
        T_{\hat\phi}H(\xi) &=& \int d^3 x\left( (\pi^{ab}-\frac{g^{ab}}{3}\sqrt {g}\langle \pi\rangle (1-e^{6\hat\phi}))\mathcal{L}_\xi g_{ab}+4(\pi-\langle\pi\rangle\sqrt g(1-e^{6\hat\phi}))\mathcal{L}_\xi\phi\right)\\
        &=& \int d^3x \left(\pi^{ab}\mathcal{L}_\xi g_{ab}-\frac{2}{3}\sqrt {g}\langle \pi\rangle (1-e^{6\hat\phi})\xi^a_{~;a}+\pi_\phi\mathcal{L}_\xi\phi+4e^{6\hat\phi}\sqrt {g}\langle \pi\rangle \mathcal{L}_\xi\phi\right)\\
        &\dot=& \int d^3 x \left(\pi^{ab}\mathcal{L}_\xi g_{ab}+\pi_\phi\mathcal{L}_\xi\phi\right),
        \end{eqnarray*}
         where we have used integration by parts and the fact that the charge constraint $\mathcal{Q}$ vanishes on the image of $T_\phi$ strongly. Thus the constraint (in smeared density form) $\pi^{ab}\mathcal{L}_\xi g_{ab}+\pi_\phi\mathcal{L}_\xi\phi$ explicitly generates diffeomorphisms in extended phase space. Its reduction to the partial theories is then $\pi^{ab}\mathcal{L}_\xi g_{ab}$, thus also generating the appropriate diffeomorphisms.
    \item $N_0$ is given by the equation
    \begin{equation}\label{lapse}(\nabla^2+\frac{1}{4}\langle\pi\rangle^2-R)N_0=\langle \Delta N\rangle,\end{equation}
    where $\Delta$ is the differential operator appearing on the lhs. One should note though that  $\pi-\langle\pi\rangle\sqrt g=0$ is a first class constraint \emph{only} in shape dynamics, thus one cannot input this constraint into equation \eqref{lapse}. The reason is that only $T_{\phi_0}N_0$ is used in shape dynamics (the solution to the $T_\phi$-transformed lapse fixing equation), and thus one would have to input said relation into that transformed equation. Likewise, in the linking theory one could input only the total charge $\mathcal{Q}$ to obtain a lapse fixing equation. However to perform the phase space reduction, one sets $\pi_\phi=0$, which reduces $\mathcal Q$ to $\pi-\langle\pi\rangle\sqrt g$ and one can thus use \eqref{lapse}. 
     Equation \eqref{lapse} is solved by $\Delta N_0=\alpha$, where $\alpha$ is any spatial constant. We fix it through the normalization condition $\langle N_0\rangle=1$, a freedom that originates from the fact that the map $\phi\mapsto \hat\phi$ is a contraction taking a general conformal factor to one that preserves the total volume. This means, that there is a unique normalized $\hat N_0[g,\pi]$ that solves the equation. Note that here $N_0$ functionally depends also on $\pi^{ab}$. 
        \item In the same way we find that given this $N_0$ we can find a purely second class set of constraints $\widetilde{T_\phi S}(x)$, which vanishes when smeared by $N_0$. We then
            have an invertible ``sub-matrix" $\{\widetilde{T_\phi S},\pi_\phi\}$. Then $\hat\phi_0$ is solved for in the same way, noticing once again that it is for $\hat\phi_0$ that we solve uniquely for, and not $\phi_0$.
            \item The leftover constraints are once more:
            \begin{equation}
  \langle T_{\phi_0}N_0\mathcal{H}\rangle,~\mathcal{D},\mathcal{H}^a.
\end{equation} The non-zero part of the (smeared) constraint algebra is given by:
\begin{eqnarray*}
\{H^a(\eta_a),H^a(\xi_a)\}&=&H^a([\xi,\eta]_a)\\
\{\langle T_{\phi_0}N_0\mathcal{H}\rangle,\mathcal{D}(\rho)\}&=&0\\
\{H(\xi),\pi(\rho))\}&=&\mathcal{D}(\mathcal{L}_\xi \rho)
\end{eqnarray*}
\end{enumerate}

\end{document}